\documentclass[11pt,dvips]{article}
\usepackage{epsfig,times} 
\usepackage{picinpar}
\usepackage{wrapfig}
\usepackage{floatflt}
\makeatletter
\renewcommand{\section}{\@startsection{section}{1}{0in}
	{0.4\baselineskip}{0.1\baselineskip}{\Large\bf}}
\renewcommand{\subsection}{\@startsection{subsection}{2}{0in}
	{0.25\baselineskip}{-\baselineskip}{\large\bf}}
\renewcommand{\subsubsection}{\@startsection{subsubsection}{3}{0in}
	{0.1\baselineskip}{-\baselineskip}{\normalsize\bf}}

\makeatother
\def\lsim{\mathrel{\hbox{\rlap{\hbox{\lower4pt\hbox{$\sim$}}}\hbox{$<$}}}}
\def\gsim{\mathrel{\hbox{\rlap{\hbox{\lower4pt\hbox{$\sim$}}}\hbox{$>$}}}}

\begin{document}

%
\thispagestyle{myheadings}

\markright{OG 4.3.28}
\begin{center}

{\LARGE \bf The Very Energetic Radiation Imaging Telescope Array System (VERITAS)}
\end{center}

\begin{center}

{\bf S.M. Bradbury$^{1}$, I.H. Bond$^{1}$, A.C. Breslin$^{2}$,
J.H. Buckley$^{3}$, D.A. Carter-Lewis$^{4}$, M. Catanese$^{4}$,
S. Criswell$^{5}$, B.L. Dingus$^{6}$, D.J. Fegan$^{2}$,
J.P. Finley$^{7}$, J. Gaidos$^{7}$, J. Grindlay$^{8}$,
A.M. Hillas$^{1}$, K. Harris$^{5}$, G. Hermann$^{9}$,
P. Kaaret$^{10}$, D. Kieda$^{6}$, J. Knapp$^{1}$, F. Krennrich$^{4}$,
S. LeBohec$^{4}$, R.W. Lessard$^{7}$, J. Lloyd-Evans$^{1}$,
B. McKernan$^{2}$, D. M\"uller$^{9}$, R. Ong$^{9}$,
J.J. Quenby$^{11}$, J. Quinn$^{2}$, G. Rochester$^{11}$,
H.J. Rose$^{1}$, M. Salamon$^{6}$, G.H. Sembroski$^{7}$,
T. Sumner$^{11}$, S. Swordy$^{9}$, V.V. Vassiliev$^{5}$,
T.C. Weekes$^{5}$}\\ {\it $^{1}$Dept. of Physics \& Astronomy,
University of Leeds, Leeds, LS2 9JT, U.K.\\
$^{2}$National University of Ireland, Dublin, Ireland\\
$^{3}$Washington University, St Louis, MO 63130, U.S.A.\\
$^{4}$Iowa State University, Ames, IA 50011, U.S.A.\\
$^{5}$FLWO, SAO, Amado, AZ 85645-0097, U.S.A.\\
$^{6}$University of Utah, Salt Lake City, UT 84112, U.S.A.\\
$^{7}$Purdue University, West Lafayette, IN 47907, U.S.A.\\
$^{8}$Harvard College Observatory, Cambridge, MA 02138, U.S.A.\\
$^{9}$University of Chicago, Chicago, IL 60637, U.S.A.\\
$^{10}$Smithsonian Astrophysical Observatory, Cambridge, MA 02138, U.S.A.\\
$^{11}$Imperial College, London, SW7 2BZ, U.K. }
\end{center}

\begin{center}
{\large \bf Abstract\\}
\end{center}
\vspace{-0.5ex}

We give an overview of the current status and scientific goals of
VERITAS, a proposed hexagonal array of seven 10 m aperture imaging
Cherenkov telescopes.  The selected site is Montosa Canyon (1390 m
a.s.l.) at the Whipple Observatory, Arizona. Each telescope, of 12 m
focal length, will initially be equipped with a 499 element
photomultiplier camera covering a 3.5 degree field of view. A central
station will initiate the readout of 500 MHz FADCs upon receipt of
multiple telescope triggers.  The minimum detectable flux sensitivity
will be 0.5\% of the Crab Nebula flux at 200 GeV. Detailed simulations
of the array's performance are presented elsewhere at this meeting.
VERITAS will operate primarily as a $\gamma$-ray observatory in the 50
GeV to 50 TeV range for the study of active galaxies, supernova
remnants, pulsars and gamma ray bursts.

%

\vspace{1ex}

\section{Introduction:}
\label{intro.sec}
The history and present status of the atmospheric Cherenkov imaging
technique has been reviewed by Ong (1998). Its great contribution to
ground-based VHE astronomy has led to burgeoning designs for ``next
generation'' instruments of increased collection area and complexity,
one of which is the Very Energetic Radiation Imaging Telescope Array
System (VERITAS), first proposed to the Smithsonian Institution in
1996. Our design study has culminated in a detailed proposal by Weekes
et al. (1999) to build an array of 10m aperture Cherenkov telescopes
in Montosa Canyon in southern Arizona.  This is a topographically flat,
dark site, at 1390 m a.s.l., close to the Whipple Observatory which
will provide the necessary infrastructure.

VERITAS will consist of six telescopes located at the corners of a
hexagon of side 80\,m with a seventh at the centre. The telescopes'
structure will be similar to the design of the Whipple 10m reflector,
which has withstood mountain conditions for over thirty years. By
employing largely existing technology in the first instance and
stereoscopic imaging, the power of which has recently been
demonstrated by HEGRA (Daum et al., 1997), we expect VERITAS to achieve the
following:
\begin{verse}
1) {\it Effective area}: $\gsim$0.1\,km$^2$ at 1\,TeV.\\
2) {\it Effective energy threshold}: $\lsim$100\,GeV with significant
sensitivity at 50\,GeV.\\
3) {\it Energy resolution}: 10\% - 15\% for events in the range 0.2 to 10\,TeV.\\
4) {\it Angular Resolution}: $\lsim$0.05$^\circ$ for
individual photons; source location to better than 0.005$^\circ$.\\
\end{verse}

\begin{figwindow}[1,r,%
{\mbox{\epsfig{file=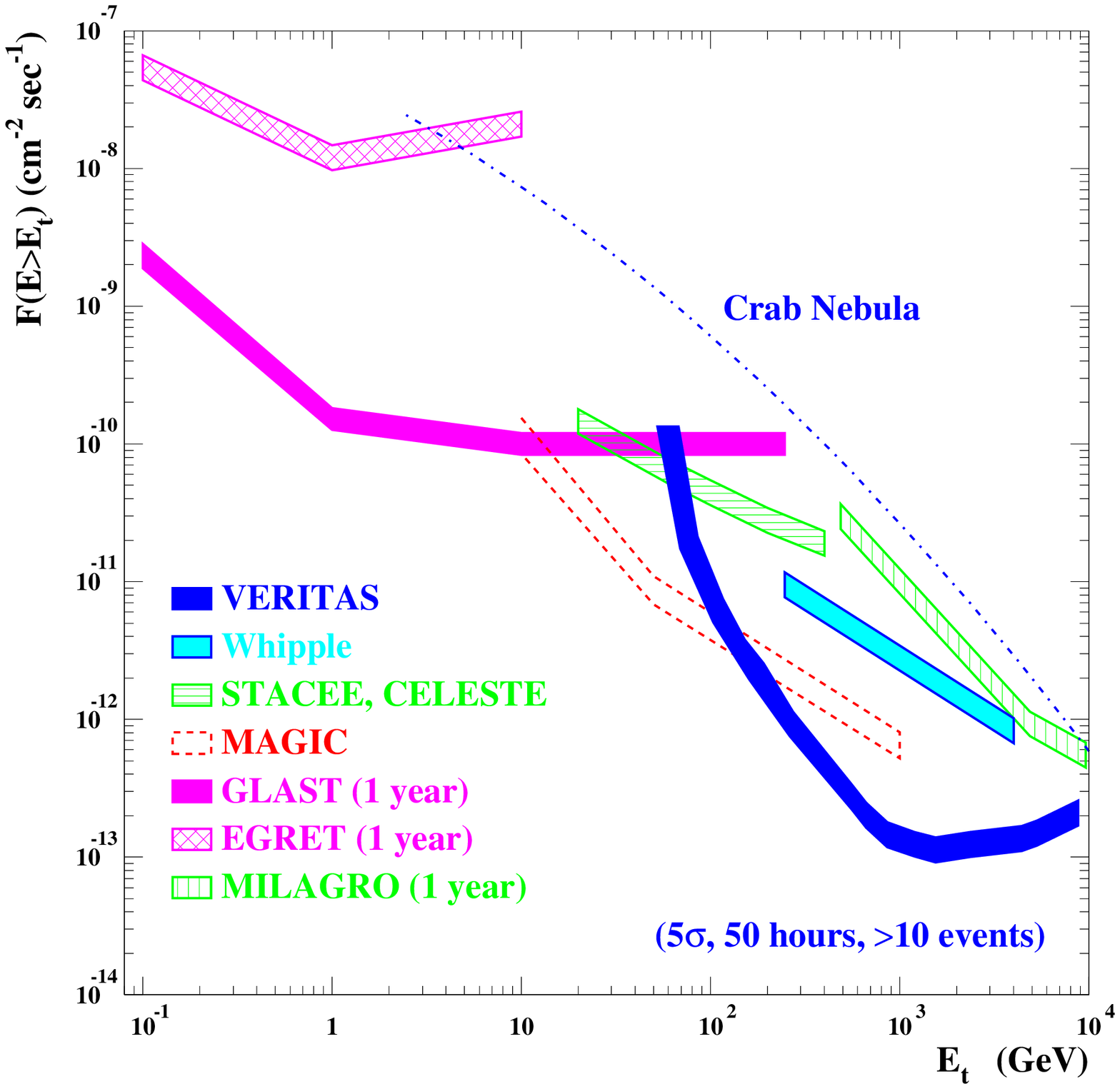,width=4in}}},%
VERITAS' sensitivity to point-like sources as compared to those of Whipple,
MAGIC, CELESTE/STACEE, GLAST and MILAGRO (Weekes et al. (1999) and references
therein).]
The performance of VERITAS is perhaps best summarised by its flux
sensitivity versus energy, shown in Figure 1 for an object of spectrum
dN/dE $\propto$ E$^{-2.5}$. Here we define a minimum detectable flux
for VERITAS as that giving a 5$\sigma$ excess of $\gamma$-rays above
background (or 10 photons where the statistics become Poissonian).  We
expect to detect sources which emit at levels of 0.5\% of the Crab
Nebula at energies of 200\,GeV in 50 hours of observation.  VERITAS,
together with the southern hemisphere Cherenkov telescope arrays HESS
and CANGAROO \mbox{-III}, will obtain high sensitivity in the 100\,GeV to
10\,TeV range between space-borne instruments and air shower
arrays. If sources of UHE cosmic rays are discovered Cherenkov
telescopes may further localise and identify them. Also, the
MILAGRO wide-field water Cherenkov detector will be sensitive to
transient sources, which, once detected can be studied in more detail
by VERITAS.

This report highlights the physics goals and some technical aspects of
VERITAS emerging from the core proposal/design study. Monte Carlo
simulations of the array's performance are presented by Vassiliev et
al. (1999).

\end{figwindow}

\section{Physics Highlights:}
\label{Phys.sec}
At a capital cost of $\sim$\$16\,M, less than 10\% of that of the
Gamma-ray Large Area Space Telescope, VERITAS will be an excellent
investment in terms of scientific return.

The large effective area of VERITAS ($\gsim 2.5\times 10^4$\,m$^2$ at
200\,GeV) will allow accurate measurements of extremely short
variations in $\gamma$-ray flux.  We illustrate this in Figure 2.  For
dense temporal coverage, VERITAS can be divided into dedicated
sub-arrays, with one sub-array observing a single object, e.g. throughout a
multi-wavelength campaign. Thus, we minimise the impact on other
scientific programmes.

\begin{figure}[t]
\centerline{\epsfig{file=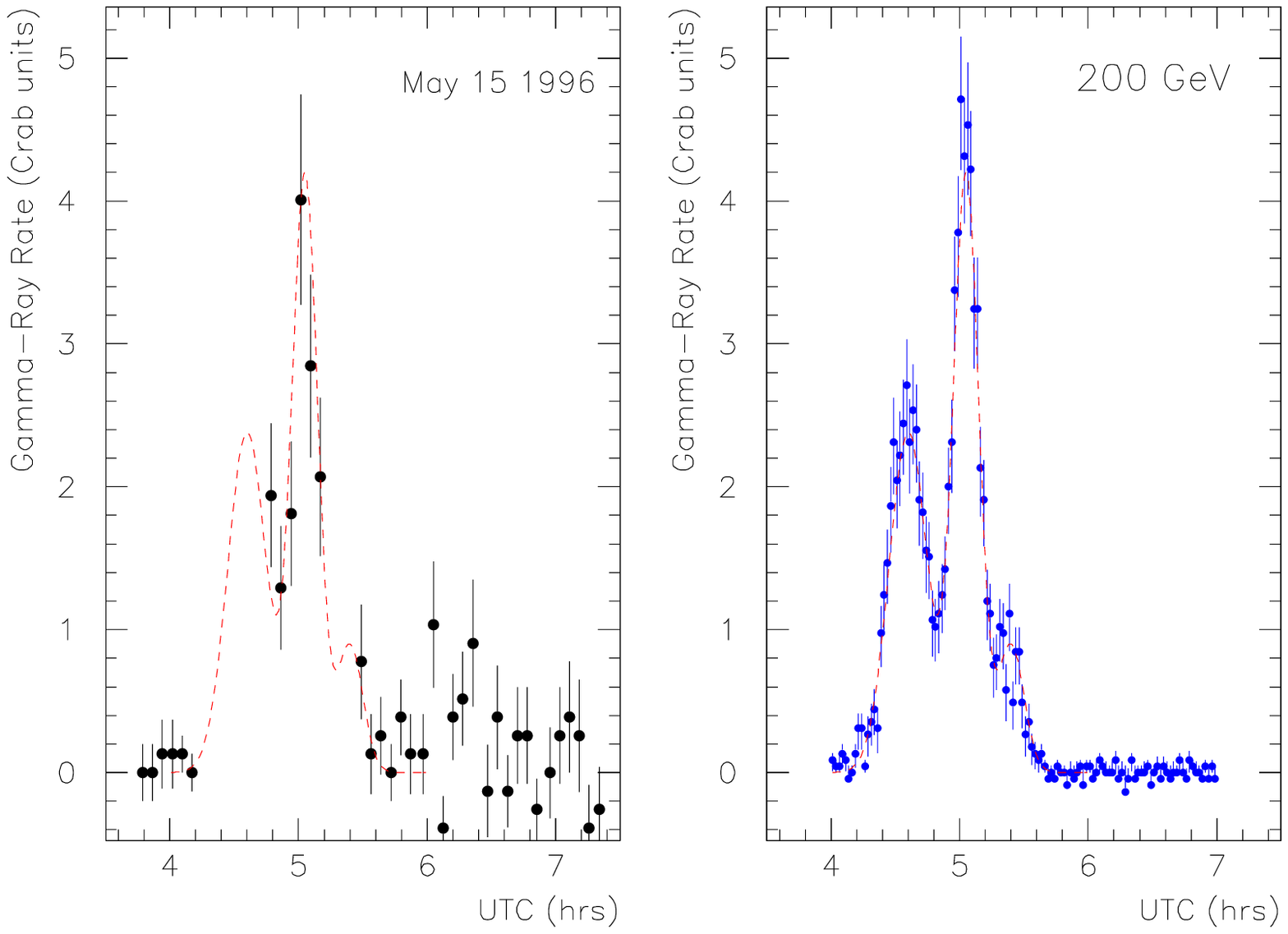,width=4.5in}}
\caption{{\it Left:} Observation of a VHE flare from Mrk 421
(Gaidos et al. 1996).  The dashed curve is a possible intrinsic flux
variation consistent with this data.  {\it Right:} Simulated
response of VERITAS to such a flare above 200\,GeV (GLAST would detect
$\sim$3 photons above 1\,GeV from a flare of this duration and power).
\label{main-960515-fig}
}
\end{figure}

\subsection{Extragalactic Astrophysics:}
\label{Exgal.sec}

We estimate that VERITAS will detect $\sim$11 of the active galactic
nuclei identified at EGRET energies, or more if they are observed in
high emission states.  VERITAS should also detect $\ge$ 30 X-ray
selected BL Lac objects (based on the spectra of Mrk 421 and Mrk 501)
and possibly the ``Extreme BL Lacs'' hypothesised by Ghisellini
(1999). We hope to distinguish intrinsic spectral features from those
due to pair-production on the IR background.  A large sample of energy
spectra for a single class of blazar will dramatically improve our
estimate of the background IR density.

Gamma ray bursts should be visible out to z$\approx$1 or more at
energies $\le$100 GeV.  VERITAS' rapid slew speed, good angular
resolution and field of view (up to 10$^\circ$ with offset pointing of
individual telescopes to cover the position error box) make it
excellent for counterpart searches. Attenuation at high
energies from interaction with background IR fields could provide
distance bounds, if source energetics are known.

\subsection{Galactic Astrophysics:}
\label{Gal.sec}

For a typical supernova remnant (SNR) luminosity and angular extent,
VERITAS should be able to detect such objects within 4\,kpc of Earth
according to the Drury, Aharonian \& V\"olk (1994) model of
$\gamma$-ray production by hadronic interactions ($\sim$ 20 shell-type
SNRs are known to lie within that range). As regards plerions and
pulsars, VERITAS should be sufficiently sensitive to detect Crab-like
objects anywhere within the Galaxy if their declination is
$>-28^\circ$. The detection of pulsed $\gamma$-rays above 50\,GeV
might be decisive in favor of the outer gap pulsar model over the
polar cap model, as the latter predicts a sharp spectral cut-off at
low energies.

For an 80-night survey of the galactic plane region $0^\circ < l <
85^\circ$, VERITAS would be sensitive to fluxes down to $\sim$0.02
Crab above 300\,GeV. This part of the sky includes 19 young, energetic
pulsars and 12 hitherto unidentified EGRET sources.  In addition,
estimates of the annihilation line flux for neutralinos at the
galactic center by Bergstr\"om, Ullio \& Buckley (1998) predict a
signal potentially detectable by VERITAS.

\section{Design:}
\label{design.sec}
\subsection{Telescope Structure:}
The telescopes will be constructed following the Davies-Cotton
reflector design with spherical and identical facet mirrors of pyrex
glass (slumped and polished, aluminised and anodised) to provide the
optimum combination of optical quality and cost effectiveness.  A
study is underway to design a stress-free mounting scheme for the
hexagonal (60\,cm flat to flat) mirrors incorporating remotely
controlled motorized alignment.  The time spread in light across the
proposed $f$/1.2 reflector is only 3-4\,ns and 100\% of the light from
a point source is captured by a 0.125$^\circ$ pixel out to 1$^\circ$
from the optical axis, decreasing to 72\% at 2$^\circ$. It is then
possible to match the inherent angular fluctuations in the shower images with
a camera that has a reasonable number of pixels (499) and a field of
view (3.5$^\circ$) which is large enough to conduct surveys and
observe extended ($\le1^\circ$) sources efficiently. For the optical support
structure of trussed steel on a commercially available pedestal,
effects due to gravitational slumping during slewing will be less than
2.2 mm on the camera face. Slew speed can be as high as 1$^\circ$ per
second on both axes.

\subsection{Electronics:}
\label{elec.sec}
At present, the need for a low noise, high gain ($>$ 10$^6$), photon
counting detector, with risetimes of less than a few ns is satisfied
only by photomultipliers (PMTs).  The Hamamatsu R7056, with a bialkali
photocathode, UV glass window and 25\,mm active diameter meets our
requirements. The collection efficiency will be increased by Winston cones.
The spacing between the PMTs will correspond to a focal plane angular 
distance of $0.15^\circ$. We plan to use a modular high voltage supply
(e.g. LeCroy 1458) where each PMT has a separately programmable high 
voltage, supplied from a system crate located at the base of the telescope.

Each PMT signal will be taken via a linear amplifier in the focus box
to a custom CFD/scaler module. The targeted gain of the PMT plus
amplifier (based on a standard 1GHz bandwidth integrated circuit chip,
AD8009) provides a signal level of $\sim$2\,mV/photoelectron.  Optical
fibre signal lines may be an attractive alternative to RG58 coaxial
cabling allowing the CFD and downstream electronics to be located in
the central control building.  A prototype multi-channel analog
optical fiber system is now being installed on the Whipple 10\,m
telescope.
 
The CFD/scaler board, incorporating an adjustable channel by channel
delay, will provide an analog fanout of the PMT signal to an FADC
system.  A prototype 500\,MHz FADC system, each channel using a
commercially available 8-bit FADC integrated circuit, has been
developed and tested successfully on the Whipple telescope (Buckley et
al. 1999).

VERITAS will operate at a minimal threshold by requiring a time
coincidence of adjacent pixel signals to form a single telescope
trigger and $>$ n coincident telescope triggers to initiate data
recording. For example, at a threshold of 5 photoelectrons the CFD
trigger rate of a single pixel will be $\sim$300\,kHz. A telescope
trigger will then require a coincidence of $\ge$ 3 {\it neighbouring}
pixel signals. This topological trigger will be similar to that used
on the Whipple telescope (e.g. Bradbury et al. 1999).  Telescope
triggers will be received at a central station where they can be used
to immediately initiate a telescope readout, or delayed to account for
orientation of the shower front (e.g. by a CAEN V486 digital delay)
and combined in a more complex trigger requirement. For example, if an
array trigger requires that 3 of 7 telescopes trigger within a 40\,ns
coincidence window then the {\it accidental} array trigger rate is $<$
1Hz at the 5 photoelectron trigger threshold.

The acquisition system architecture will be based largely on a fast
VME backplane and distributed computation by local CPUs running a
real-time operating system. For an array trigger rate of $\sim$1\,kHz
each telescope is expected to have an average data flow rate of
$\sim$800\,kbyte/s, resulting in a 200\,kbyte/s rate on any VME
backplane. Each crate controller will be connected to a local
workstation which in turn will communicate with a central CPU. The
central CPU will perform control and quicklook functions, data
integration and compression.  Telescope guidance and high voltage
control will be performed by inexpensive Pentium PCs.

\section{Conclusion:}
\label{Con.sec}
Our aim is to commission the first VERITAS telescope in 2001. The 10m
Whipple telescope nearby will operate throughout the construction
phase and serve as a test-bed for innovative technologies. By staged
construction of the array over a four year period, we expect to
maintain an astronomical facility at all times. VERITAS is expected to
be in routine operation prior to the launch of GLAST in 2005.
\vspace{1ex}
\begin{center}
{\Large\bf References}
\end{center}
%
Bergstr\"om, L., Ullio, P. \& Buckley, J.H. 1998, Astrop. Phys., 9, 137\\
Bradbury, S.M. et al. 1999, Proc. 26th ICRC (Salt Lake City, 1999), OG 4.3.21\\
Buckley, J.H. et al. 1999, Proc. 26th ICRC (Salt Lake City, 1999), OG 4.3.22\\
Daum, A. et al. 1997, Astrop. Phys., 8, 1\\
Drury, L.O'C., Aharonian, F.A. \& V\"olk, H.J. 1994, A\&A, 287, 959\\
Gaidos, J.A. et al. 1996, Nature, 383, 319\\
Ghisellini, G. 1999, in ``TeV Astrophysics of Extragalactic Sources'', 
Astrop. Phys., eds. M. Catanese \& T.C. Weekes, in press\\
Hurley, K. 1994, Nature, 372, 652\\
Ong, R.A. 1998, Physics Reports, 305, 93\\
Vassiliev, V.V. et al. 1999, Proc. 26th ICRC (Salt Lake City, 1999), OG 4.3.35\\
Weekes, T.C. et al. 1999, VERITAS, proposal to SAGENAP\\

\end{document}